\documentclass{article}
\begin{document}
\title{$\kappa$--deformed Wigner construction of
 relativistic   wave functions and  free fields
 on $\kappa$-Minkowski space\thanks{
This paper is    supported by KBN grant 5PO3B05620.}}

\author{
Piotr Kosi\'nski
\\
Department of Theoretical Physics II \\
University of {\L}\'od\'z \\
Pomorska 149/153, 90 - 236 {\L}\'{o}d\'{z}, Poland
\\
Jerzy Lukierski
\\
Institute for
Theoretical Physics, University of Wroc{\l}aw,\\
pl. Maxa Borna 9, 50-205 Wroc{\l}aw, Poland  
\\
 and Pawe{\l} Ma\'{s}lanka
\\
Department of Theoretical Physics II \\            
University of {\L}\'od\'z \\                        
Pomorska 149/153, 90 - 236 {\L}\'{o}d\'{z}, Poland
}

\date{}		
\maketitle
\begin{abstract}
We describe the extension of the Wigner`s infinite-dimensional
unitary representations of Poincar\'{e} group to the case of
$\kappa$-deformed  Poincar\'{e} group. We show that the
corresponding coordinate wave functions on noncommutative
space-time are described by free field equations on
$\kappa$-deformed Minkowski space.
 The cases of Klein--Gordon, Dirac, Proca and Maxwell fields are considered.
 Finally some aspects of second quantization are also discussed.
\end{abstract}

\section{Introduction}

Recently, there has been much interest in field theory on
noncommutative space-time  see e.g. [1-3]. Most authors deal with
the simplest case when coordinates commute to a c-number; this
implies that
 classical Poincar\'{e} symmetries are not modified,
 Lorentz symmetry however is explicitly broken. A slightly
more complicated structure is obtained if one assumes that the
 noncommutative space--time coordinates form a Lie algebra,
  with ``quantum" time
 and commutative space   coordinates [4]. It appears then that,
although
more complicated, such quantum Minkowski space admits
10--parameter
symmetry  with  noncommuting quantum group parameters.
Such a symmetry  is given by particular   deformation of Poincar\'{e} group called $\kappa$%
-Poincar\'{e} quantum group $P_{\kappa}$ with classical subgroup
of
 nonrelativistic $O(3)$ rotations [5]. Due to the fact
that in such a case the
deformation parameter is  dimensionful
 and can be related to Planck length ($l_{p} \simeq 10^{-33}$cm)
 one can speculate that
$\kappa$%
-Poincar\'{e}  symmetry could describe the space-time structure
at Planck scale  (see e.g. [6]). We suppose, therefore, that it
  is interesting to elaborate in
more detail the consequences of the assumption  that
$P_{\kappa}$\ is a generalization of standard space-time symmetry
at very high energies. First steps towards the construction of
quantum field theory based on $P_{\kappa}$\ were made recently in
[7]. This paper has as its aim the group--theoretic derivation of
$\kappa$-deformed
 free field theories.
\newline
We would like to discuss here a particular  aspect of
$P_{\kappa}$\ symmetry. In standard field  theory one starts with
free fields which provide a building block for the construction of
Lagrangeans,  defining propagators etc.; they are basic elements
of the theory on perturbative level [5]. In order to define the
free fields one has to solve the basically group-theoretical
problem: to construct the intertwinners between the
representations of Poincar\'{e} group acting on fields  and
states. These intertwinners are simply the wave functions of
particles of definite mass and spin and are related with Wigner
theory of unitary representations of Poincar\'{e} group [8]. It is
customary to define them through the appropriate wave equations
describing free fields.
\newline Below we
solve the same problem in the          $\kappa$-deformed case. The
solution appears to be quite simple and, basically, reduces to
the one for the nondeformed ($\kappa=\infty$) case. It is
important to stress that our problem is of group-theoretical
origin. In such a case we speak about coordinate wave functions
rather than quantum fields. Here we shall consider
$\kappa$-deformed classical fields, with deformation originating
from $\kappa$-deformation of Minkowski space.
 To obtain quantum
$\kappa$--deformed  fields one has only to replace their Fourier
transforms describing momentum amplitudes by relevant
creation/anihilation operators. We would like to point out that
 solving
the problems of quantum (second--quantized) $\kappa$--deformed
free fields is outside of the scope of this talk.
\newline To conclude this section let us define our basic
notions: the quantum Minkowski space $M_{\kappa}$\ and
$\kappa$-Poincar\'{e} group $P_{\kappa}$.
\newline
$M_{\kappa}$\ is a $*$-algebra generated by four Hermitian elements
$x^{\mu}$%
obeying
\begin{eqnarray}\label{wz1}
[x^{\mu},x^{\nu}]=\frac{i}{h}(\delta_{0}^{\mu}x^{\nu} -
\delta_{0}^{\nu}x^{\mu}).
\nonumber
\end{eqnarray}
We shall use in the sequel the following "normal" ordering which
gives the correspondence between $M_{\kappa}$\ and standard
Minkowski space. The product of $x^{\mu}$'s is called normally
ordered if all $x^{0}$\ factors stand leftmost; we denote normal
product by : :. To any analytic
 function $%
f(x)$\ we can ascribe a (formal) element $\hat{f}$\ of $M_{\kappa}$\
by
\begin{eqnarray}\label{wz2}
\hat{f}=: f(x):  \nonumber
\end{eqnarray}

The quantum $\kappa$-Poincar\'{e} group $P_{\kappa}$\ [2] is a
$*$-Hopf algebra generated by hermitean elements
$\Lambda^{\mu}_{\;\;\nu}$\ and $a^{\mu}$\ obeying
\begin{eqnarray}\label{wz3}
&&g_{\mu \nu}\Lambda_{\;\;\alpha}^{\mu}\Lambda_{\;\;\beta}^{\nu}=g_{\alpha
\beta}I
\nonumber \\
&&[a^{\mu},a^{\nu}]=\frac{i}{\kappa}(\delta_{0}^{\mu}a^{\nu}-
\delta_{0}^{%
\nu}a^{\mu})  \nonumber \\
&&[\Lambda_{\;\;\nu}^{\mu},a^{\rho}]=-\frac{i}{\kappa}((\Lambda_{\;\;0}^{\mu}-
\delta_{0}^{\mu})\Lambda^{\rho}_{\;\;\nu}
+(\Lambda_{\;\;\nu}^{0}-\delta_{\nu}^{0})g^{\mu \rho})  \nonumber \\
&&[\Lambda_{\;\;\nu}^{\mu},\Lambda_{\;\;\beta}^{\alpha}]=0  \nonumber \\
&&\Delta(\Lambda_{\;\;\nu}^{\mu})=\Lambda_{\;\;\alpha}^{\mu}\otimes
\Lambda_{\;\;\nu}^{\alpha}  \nonumber \\
&&\Delta(a^{\mu})=\Lambda_{\;\;\nu}^{\mu}\otimes a^{\nu}+a^{\mu}\otimes I
\nonumber \\
&&S(\Lambda_{\;\;\nu}^{\mu})=\Lambda_{\nu}^{\;\;\mu}  \nonumber \\
&&S(a^{\mu})=-\Lambda_{\;\;\nu}^{\mu}a^{\nu}  \nonumber \\
&&\varepsilon(\Lambda^{\mu}_{\;\;\nu})=\delta^{\mu}_{\nu}I  \nonumber \\
&&\varepsilon(a^{\mu})=0\;\;\;;  \nonumber
\end{eqnarray}

here $g_{\mu \nu}=diag(+---)$\ is numerical metric tensor.

\section{Unitary representations of $\protect\kappa$-Poincar\'{e}
group}

Let us recall the construction of induced representations [9]. One
starts with some subgroup $H \subset G$\ of the group under
consideration. The relevant Hilbert space is the space of  square
integrable functions on $G$\ taking their values in the vector
space carrying some representation
of $H$. %
 The group action is defined to be, say, right action: $%
g_{0}:f(g)\rightarrow f(gg_{0})$.  The essence of the method is
the selection of invariant subspace by imposing coequivariance
condition;  in many cases the invariant subspace obtained in this
way carries an irreducible representation  of $G$.  The more
explicit characterization of the representation is achieved by
solving explicitly the  coequivariance condition which gives rise
to the description of the representation in terms of Hilbert
space of functions defined on the right coset space $H \backslash
G$.  This is especially effective for the case of semidirect
products in which one factor is abelian. In particular, in the
case of Poincar\'{e} group we obtain Wigner's
construction.\newline The whole construction can be generalized
in a rather straightforward way to the $\kappa$--deformed  case
[10]. The actual construction for the
case of $\kappa$%
-Poincar\'{e} group has been given in Ref. [11]. The results read:
\newline (a) \underline{the massive case} \newline The carrier
space is the space of square-integrable (with respect to the
invariant measure $d\vec{q}/{q}^{0}$) functions on hyperboloid
$(q^{0})^{2}-%
\vec{q}^{2}=m^{2}, \;q^{0}>0$,  taking their values in the
representation space of spin $s$ representation of rotation group
($s$\ should be integer as we are dealing with standard
``vectorial"  Poincar\'{e} group).\newline The right coaction of
$P_{\kappa}$\ is then given by
\begin{eqnarray}\label{wz4}
&&  f(q,\sigma)\rightarrow
\sum_{\sigma '}D_{\sigma \sigma '}^{(s)}
(R(q\otimes I,I\otimes \Lambda))
\cr
&&
\quad
e^{-iP_{0}(q)\otimes
a^{0}}
e^{-iP_{k}(q)\otimes a^{k}}f(q_{\nu}\otimes \Lambda^{\nu}_{\;\;\mu} ,
\sigma '); \label{w1}
\end{eqnarray}
where
\begin{eqnarray}\label{wz5}
&&P_{0}(q)=\kappa
\ln({\rm ch}(\frac{m}{\kappa})+\frac{q_{0}}{m}{\rm sh}(\frac{m}{\kappa}))
\label{w2} \\
\nonumber \\
&&P_{k}(q)=\frac{\kappa
{\rm sh}(\frac{m}{\kappa})q_{k}}{m {\rm ch}(\frac{m}{\kappa})%
+q_{0}{\rm sh}(\frac{m}{\kappa})}  \label{w3}
\end{eqnarray}
and $D^{(s)}(R(q\otimes I,I\otimes \Lambda))$\ is the spin $s$\
representation of classical Wigner rotation written in
appropriate tensor product form.\newline (b) \underline{the
massless case} \newline The carrier space is now the space of
$\bf{C}$-valued functions on upper light cone,
$(q^{0})^{2}-\vec{q}^{2} =0, \;q^{0}>0$, square-integrable with
respect to the same measure $d^{3}\vec{q}/q^{0}$. The right
coaction of $P_{\kappa} $\ is now given by
\begin{eqnarray}\label{wz6}
&& f(q)\rightarrow
 e^{i\lambda \Theta (q\otimes I,I\otimes
\Lambda)}e^{-iP_{0}(q)\otimes a^{0}}
\cr
&&
\quad
e^{-iP_{k}(q)\otimes a^{k}}
f(q_{\nu}
\otimes\Lambda_{\;\;\mu}^{\nu});  \label{w4}
\end{eqnarray}
$\lambda$\ is the (integer) helicity, $\Theta (q\otimes I,I\otimes
\Lambda)$%
-the classical rotation angle of the little group $E(2)$\ while
\begin{eqnarray}\label{wz7}
&&P_{0}(q)=\kappa \ln (\frac{q_{0}}{k}(e^{\frac{k}{\kappa}}-1)+1)
\label{w5}
\\
\nonumber \\
&&P_{k}(q)= \frac{\kappa(e^{\frac{k}{\kappa}}-
1)q_{k}}{q_{0}(e^{\frac{k}{%
\kappa}}-1)+k}  \label{w6}
\end{eqnarray}
and $k$\ parametries the standard fourvector $(k,0,0,k)$\ \newline
It is easy to check [12] that the infinitesimal form of the above
representation is the representation of the $\kappa$-Poincar\'{e}
algebra in bicrossproduct  basis with classical Lorentz algebra
sector [13]. It has been shown in [11] that the $\kappa$-Poincar\'{e}
group and $\kappa$-Poincar\'{e} algebra are related by Hopf algebra
duality; the explicit form of duality relations in M-R basis is
given by
\begin{eqnarray}\label{wz8}
&&<P_{\mu},:f(a):>=i \frac{\partial f}
{\partial a^{\mu}}\mid_{a=0} \nonumber \\
&&<M_{\mu \nu},g(\Lambda )>=
i(\frac{\partial g}{\partial \Lambda^{\mu \nu}}-\frac{\partial g}
{\partial \Lambda^{\nu \mu}})\mid_{\Lambda = I} \label{w77}
\end{eqnarray}
for translations and Lorentz transformations, respectively,
extended suitably with the help of bicrossproduct structure [14].
The infinitesimal generators are then defined as follows: denote
by $\varrho_{R}$\ the right coaction (\ref{w1}); for any element
X of $\kappa$-Poincar\'{e} algebra we define
\begin{eqnarray}\label{wz9}
\hat{X}f\equiv \sum_{\alpha}<X,\varphi_{\alpha}>
f_{\alpha} \label{w88}
\end{eqnarray}
where $\varrho_{R}(f)=\sum_{\alpha}
f_{\alpha} \otimes \varphi_{\alpha}$.
 Simple calculation
gives
\begin{eqnarray}\label{wz10}
&&\hat{P}_{0}= \kappa \,
 \ln({\rm ch}(\frac{m}{\kappa})+\frac{q_{0}}{m}
 {\rm sh}(\frac{m}{\kappa})) \nonumber \\
&&\hat{P}_{k}= \frac{\kappa
{\rm sh}(\frac{m}{\kappa})q_{k}}{m\, {\rm ch}(\frac{m}
{\kappa})+q_{0}{\rm sh}(\frac{m}{\kappa})} \nonumber \\
&&\hat{M}_{ij}=i(q_{i}\frac{\partial}
{\partial q^{j}}-q_{j}\frac{\partial}{\partial q^{i}})+
\varepsilon_{ijk}S_{k} \label{w109}\\
&&\hat{M}_{i0}=-iq_{0}\frac{\partial }
{\partial q^{i}}+\varepsilon_{ijk}
\frac{q_{j}S_{k}}{q_{0}+m};\nonumber
\end{eqnarray}
here $\{ S_{k} \}$\ (spin matrices)
 span the representation of the little group. Similar results
are obtained in the case of massless representation, Eq. (\ref{w4}).

 In the bicrossproduct  basis the
Lorentz algebra acts nonlinearly in momentum space. This action
can be linearized (providing thereby the equivalence of the
algebraic sector with that of the classical Poincar\'{e} algebra)
by the following deformation map [15,16]
\begin{eqnarray}\label{wz11}
&&P_{0}(q)=\kappa \ln(\frac{q_{0}+C}{C-A})  \label{w7} \\  
\nonumber \\
&&P_{k}(q)= \frac{\kappa q_{k}}{q_{0}+C}  \label{w8}       
\end{eqnarray}
where $A$\ and $C$\
are functions of $m^{2}=q^{2} = q^{2}_{0} - q^{2}_{k}$
 \ obeying $%
A^{2}-2AC+m^{2}=0$.
The expressions appearing on the right hand side
of
Eqs. (\ref{w2}), (\ref{w3}) provide a special case  of deformation
map with $%
A=m(cth (\frac{m}{\kappa})-\frac{1}{sh(\frac{m}{\kappa})}),\;
C=mcth(\frac{m%
}{\kappa})$\ (those given by Eqs. (\ref{w5}), (\ref{w6}) correspond
to $%
A=0,\;C=\frac{k}{e^{\frac{k}{\kappa}}-1})$.  Using the properties of
deformation map (\ref{w7})
 and (\ref{w8}) one can easily show that by
replacing $P_{\mu}(q)$\ given by Eqs.
(\ref{w2}),(\ref{w3}),(\ref{w5}) and (%
\ref{w6})  by more general ones (\ref{w7}), (\ref{w8}) one obtains
equivalent representations. More precisely, for two
representations, given by Eqs. (\ref{w1})(or(\ref{w4})) and
(\ref{w7}), (\ref{w8}),   determined by the functions $C$\ and
$C^{\prime}$, which depends on  the mass  $m^{2} = q^{2}$  and
deformation parameter $\kappa$,  this equivalence is given by
\begin{eqnarray}\label{wz12}
f^{\prime}(q,\sigma)=(\frac{C^{\prime}}{C})f(\frac{C}{C^{\prime}}q,
\sigma)
\end{eqnarray}

Because
\begin{equation}\label{luma13} 
  P_{\mu}\left( {C\over C^{\prime}} \, q_{\mu},
  A^{\prime}, C^{\prime} \right)
  = P_{\mu} \left( q_{\mu}, A,C\right)
\end{equation}
the inverse formula to
 (\ref{w7}--\ref{w8})
 satisfies the relation
\begin{equation}\label{luma14} 
  q^{\prime}_{\mu} = q_{\mu} \left( P,A^{\prime},C^{\prime}\right)
  =  {C^{\prime}\over C} q_{\mu}
  \left( P, A, C \right)
\end{equation}

It follows from (14)
  that replacement $C \to C^{\prime}$ implies the
change of the mass parameter $q^{2}_{\mu} = m^{2} \to q^{\prime
2}_{\mu} =  {C^{\prime 2}\over C^{2}} \, m^{2} = m^{\prime 2}.$
Therefore, by a particular choice of the function $C$ one can relate the
deformed fourmomenta described by deformed mass shell condition
 (parametrized by  $\kappa$-Poincar\'{e}-invariant mass $M$)
\begin{eqnarray}\label{luma14b}  
  \left( 2 \kappa \sinh \, {P_{0}\over 2 \kappa} \right)^{2}
  - e^{P_{0}\over \kappa} \, p^{2} = M^{2}
\end{eqnarray}
to the  fourmomenta $q_{\mu}$, obeying on classical relativistic
mass shell condition $q_{0}^{2} - \overrightarrow{q}^{2}
 = m^{2}$ with
arbitrary choice of positive mass parameter $m$.

To deal with halfinteger spin case one has only to replace
$\kappa$-Poincar\'{e} group by $\kappa$--deformed  counterpart of
$ISL(2,C)$\ group described in [17]. The whole construction can be
repeated  almost without modifications; one has only to replace
Wigner rotation by its  classical $SL(2,C)$\ element so that
$D^{(s)} $\ becomes a function of $q_{\mu}\otimes I$\ and
$I\otimes A, A\in
SL(2,C).$%
\

\section{Covariant wave functions}

In the standard case (i.e. with nondeformed symmetry) the building
blocks for the  construction of local interactions are the
covariant fields linear in creation/anihilation operators [8].
They are obtained by solving the following problem: given
particles of a definite mass and spin we know the transformation
properties of the relevant momentum amplitudes, i.e.  the wave
functions in momentum space or, after second quantization,
momentum space creation/anihilation operators. In the following
we shall construct the appropriate coordinate space wave functions
carrying the same amplitudes and transforming according to a
given representation of  Lorentz group
\begin{eqnarray}\label{wz16}
\Phi_{l}(x)\rightarrow
D_{ll^{\prime}}(\Lambda)\Phi_{l^{\prime}}(\Lambda^{-1}(x-a))
\label{w9} 
\end{eqnarray}
Quantum fields are then obtained by replacing momentum space
amplitudes by creation or anihilation operators.\newline We
attempt here to solve the same problem in $\kappa$-deformed case.
Due to the rather mild character of $\kappa$-deformation as seen
in  the form of unitary representations given in previous section,
it is not surprising that this
problem can be solved in a quite straightforward way. Let us  recall
that $%
P_{\kappa}$\ acts covariantly
on $M_{\kappa}$\ from the left as given
by:
\renewcommand{\theequation}{17\alph{equation}}
\setcounter{equation}{0}
\begin{eqnarray}\label{wz17}
x^{\mu}\rightarrow \Lambda^{\mu}_{\;\;\nu}\otimes
x^{\nu}+a^{\mu}\otimes I
\label{w10}  
\end{eqnarray}

However, we prefer to consider the right coaction of
$\kappa$-Poincar\'{e} which corresponds to the classical
representations rather than
antirepresentations. It appears that, in order to get a covariant
action on $%
M_{\kappa}$, one has to consider the group $P_{-\kappa}$,  the
$\kappa$%
-Poincar\'{e} group with $\kappa$\ replaced by $-\kappa$.
Therefore, we define the action of $P_{\kappa}$\ on $M_{\kappa}$\
as follows
\begin{eqnarray}\label{wz18}
x^{\mu}\rightarrow x^{\nu}\otimes \Lambda_{\nu}^{\;\;\mu}-I\otimes
a^{\nu}\Lambda_{\nu}^{\;\;\mu}  \label{w11}  
\end{eqnarray}
As a next step we define the coordinate wave functions. Let us
select some representation $D^{(A,B)}$\ of classical Lorentz
group and let ${\bf C}^{n}, \;n=(2A+1)(2B+1)$\ be the carrier
space of this representation. The
coordinate wave function $\Phi_{l}(x)$\ is an element of $%
M_{\kappa}\otimes {\bf C}^{n}$.  It is then easy to check that the
following action of $P_{-\kappa}$\ is well defined
\renewcommand{\theequation}{\arabic{equation}}
\setcounter{equation}{17}
\begin{eqnarray}\label{wz19}
&&
\Phi_{l}(x)\rightarrow \Phi_{l^{\prime}}(x^{\nu}\otimes
\Lambda_{\nu}^{\;\;\mu}-I\otimes a^{\nu}\Lambda_{\nu}^{\;\;\mu})
\cr
&&
\qquad
(I\otimes
D_{ll^{\prime}}^{(AB)}(\Lambda))
\end{eqnarray}
The following result forms the basis for the whole subsequent
discussion. Let $\varphi (x)\in M_{\kappa}$\ be an element given
by
\begin{eqnarray}\label{wz20}
\varphi (x)= \int \frac{d^{3}\vec{q}}{q_{0}}a(q):e^{-
iP_{\mu}(q)x^{\mu}}:
\end{eqnarray}
where $a(q)$\ is a ${\bf C}$-valued function on the positive
hyperboloid $%
(q^{0})^{2}-\vec{q}^{2}=m^{2}$\ while $P_{\mu}(q)$\ is given by
(2,3).
\newline Then the following basic identity holds
\begin{eqnarray}\label{wz21}
&&\int \frac{d^{3}\vec{q}}{q_{0}}a(q)e^{-iP_{0}(q)(x^{\nu}\otimes
\Lambda_{\nu}^{\;\;0}-I\otimes a^{\nu}
\Lambda_{\nu}^{\;\;0})}
\cr
&&\quad
e^{-iP_{k}(q)(x^{\nu}\otimes \Lambda_{\nu}^{\;\;k}-I\otimes
a^{\nu}\Lambda_{\nu}^{\;\;k})}= \label{w15} \\  
&&= \int \frac{d^{3}\vec{q}}{q_{0}}e^{-iP_{0}(q)(I\otimes
a^{0})}
e^{-iP_{k}(q)(I\otimes a^{k})}
\cr
&&
\quad
a(q_{\nu}(I\otimes
\Lambda^{\nu}_{\;\;\mu}))
(:e^{-iP_{\mu}(q)x^{\mu}}:\otimes I)  \nonumber
\end{eqnarray}
The proof of the above identity is straightforward but tedious.
It can be also extended to massless case.\newline The basic
identity (\ref{wz21}) allows us to formulate the following general result.
Let us put
\begin{eqnarray}\label{wz22}
\Phi_{l}(x)\equiv \sum_{\sigma}\int \frac{d^{3}\vec{q}}{q_{0}}%
U_{l}(q,\sigma)a(q,\sigma):e^{-iP_{\mu}(q)x^{\mu}}:
\end{eqnarray}
where the wave functions $U_{l}(q,\sigma)$\ obey the intertwinning
conditions
\begin{eqnarray}\label{wz23}
D_{ll^{\prime}}^{(AB)}(\Lambda)U_{l^{\prime}}(q,\sigma)=D_{\sigma
^{\prime}\sigma}^{(S)}
(R(q,\Lambda))U_{l}(\Lambda q,\sigma ^{\prime}).
\end{eqnarray}
Then $\Phi_{l}(x)$\ transforms according to the formula
(\ref{wz19}). The
massless counterpart reads:
\begin{eqnarray}\label{wz24}
\Phi_{l}(x)=\int \frac{d^{3}\vec{q}}{q_{0}}U_{l}(q,\lambda)a(q):e^{-
iP_{%
\mu}(q)x^{\mu}}:
\end{eqnarray}
with $P_{\mu}(q)$\ given by
Eqs.
 (\ref{w5}, \ref{w6}) transforms covariantly if
\begin{eqnarray}\label{wz25}
D_{ll^{\prime}}^{(AB)}(\Lambda)U_{l^{\prime}}(q,\lambda)=e^{i\lambda
\Theta
(q,\Lambda)}U_{l}(\Lambda q,\lambda)
\end{eqnarray}
It is crucial that the conditions (\ref{wz23}) and
(\ref{wz25}) are purely
classical. This allows to write out immediately the deformed
counterpart of free field theory once the classical problem is
solved. Finally, let us remark that it is not difficult to show
that above reasoning is valid also if one uses a general
deformation map
(\ref{w7}, \ref{w8}).

\section{Free field equations}

In general there is some redundancy in the description in terms of
coordinate space wave functions. In fact, the representation
$D^{(AB)}$\
can, in the massive case, describe all particles of spin s obeying
$\mid
A-B\mid \leq s \leq A+B$. Therefore, apart from the Klein-Gordon
 (KG)  condition, $%
(q^2-m^2)U_{l}(q,\sigma)=0$\ there are other conditions of the form
\begin{eqnarray}\label{wz26}
\Pi_{ll^{\prime}}(q)U_{l^{\prime}}(q,\sigma )&=&0
\end{eqnarray}
singling out a definite spin. These conditions can be converted
into constant coefficients matrix differential equations in
coordinate space. The basic problem of the theory of wave
equations in undeformed as well as $\kappa$-deformed case  is to
find a single equation from which K-G equation as well as
additional conditions
(\ref{wz26}) follow and to derive it from some
Lagrangian. This problem has particular solutions depending on
$s$\ and $(A,B)$.  It is considered as a preliminary step towards
interacting field theory. In fact, such an equation defines
propagators which are necessary to formulate the perturbative
version of the theory.

The result of Sec.
 3 imply the following procedure.
Let us invert the relations
 (\ref{w7}, \ref{w8})
 which yield
\begin{eqnarray}\label{wz27}
&&q_{0}(P)=(C-A)e^{P_{0}/\kappa}-C  \nonumber \\
&&q_{k}(P)=(C-A)e^{P_{0}/\kappa}\frac{P_{k}}{\kappa}
\end{eqnarray}
For any $f\in M_{\kappa}$\ given by
\begin{eqnarray}\label{wz28}
f=\;:f(x):
\end{eqnarray}
we define the derivative operators as follows
\begin{eqnarray}\label{wz29}
i\hat{\partial}_{\mu}f=:q_{\mu}(i\frac{\partial}
{\partial x^{\nu}})f(x):
\end{eqnarray}
Now, let
\begin{eqnarray}\label{wz30}
F_{AB}(\partial )\Phi_{B} =0
\end{eqnarray}
be the relevant classical wave equation on commutative Minkowski
space. Its $\kappa$-deformed counterpart has the same form, only
the standard space--time derivatives $\partial_{\mu}$ are
replaced by the vector fields $\widehat{\partial}_{\mu}$ on
$\kappa$--deformed Minkowski space (see (23)).
\begin{eqnarray}\label{wz31}
F_{AB}(\hat{\partial })\Phi_{B} =0
\end{eqnarray}
This is the general prescription for constructing wave equations on
$\kappa$%
-Minkowski space.

Let us write the Fourier transform for $\Phi_{A}\subset
M_{\kappa}$ as follows (compare with (17))
\begin{equation}\label{luma29}   
  \Phi_{A}( \widehat{x}) = \int \, {d^{3}p \over \omega_{\kappa}
  (p)} \, \widetilde{a}_{A}(p) \, : \, e^{-ip_{\kappa}
  \widehat{x}^{\kappa}} \, :
\end{equation}
where
\begin{eqnarray}\label{luma30}       
  \widetilde{a}_{A}(p) &= & a_{A} \left( q(p) \right)
  \\
  \omega_{\kappa}(p) &= & q_{0} (p) \cdot \, \det \left(
  {\partial q_{i}\over \partial p_{j} } \right)
  \nonumber
\end{eqnarray}
one can replace the algebra of fields
 (\ref{wz30})
 on noncommutative
Minkowski space by homomorphic algebra of fields on commutative
Minkowski space  with the multiplication described by the
$\kappa$-deformed star product (see [7,18])
\begin{eqnarray}\label{luma31} 
&&  \Phi_{A}( \widehat{x})\cdot \Phi_{B} ( \widehat{x})
  \longleftrightarrow\varphi_{A} (y) \star \varphi_{B}(y) =
  \\
  && = \varphi_{A}(y + \xi_{1} ) \exp \, iy_{\mu}
\cr
&&    \qquad
    \Gamma^{\mu}
  \left( { \overleftarrow{\partial} \over \partial \xi_{1} },
  { \overrightarrow{\partial} \over \partial \xi_{2} }
  \right) \varphi_{B} (y + \xi_{2})\Big|_{\xi_{1}=\xi_{2}=0}
  \nonumber
\end{eqnarray}
where
\begin{eqnarray}\label{luma32} 
  \Gamma_{o}(\eta,\xi) & =& \eta^{0} +\xi^{0}
  \cr
   \Gamma^{i}(\eta,\xi) &= &
{f_{\kappa}(\eta^{0}) e^{\xi_{0}\over \kappa} \eta^{i} +
f_{\kappa} (\xi^{0})
 e^{ \eta_{0}\over \kappa } \, \xi_{i} \over
f_{\kappa}(\eta^{0} +\xi^{0} )}
\end{eqnarray}
and $f_{\kappa}(\alpha) = {\kappa \over \alpha}  (e^{\alpha \over
\kappa} - 1)$.

The field equation
 (\ref{wz31})
 on noncommutative space is  translated
into the $\kappa$-deformed
 field equation on standard Minkowski space as
follows
\begin{equation}\label{luma33} 
  F_{AB} \left(
q_{\mu} \left(
i  {\partial \over \partial  y^{\nu}}
 \right) \right) \phi_{B} (y) = 0
\end{equation}
Further we shall consider the following simple choice of the
formulae
 (\ref{wz27}), corresponding   to $C=(m^{2} +\kappa^{2})^{1\over
2}$, leading to (see e.g. [15])
\begin{eqnarray}\label{luma34} 
  q_{0}(P_{\mu})& =& \kappa \sinh {P_{0}\over \kappa} +
  {1\over 2 \kappa} \, e^{P_{0}\over \kappa}
  \overrightarrow{p}^{2}
  \cr
  q_{i}(P_{\mu})
& = & e^{P_{0}\over     \kappa}  P_{i}
 \end{eqnarray}
Below we shall consider  the $\kappa$-deformed counterpart of the
Klein--Gordon, Dirac, Proca and Maxwell equations.

(a) Klein--Gordon equation.

On $\kappa$-deformed Minkowski space the KG   field equation
looks as follows:
\begin{equation}\label{luma35}  
  \left( \widehat{\partial}_{\mu} \widehat{\partial}^{\mu} -
  m^{2}_{0} \right)
  \Phi( \widehat{x}) = 0
\end{equation}
If we observe that
\begin{equation}\label{luma36} 
  q_{\mu}q^{\mu} = {\cal M}^{2} (p)
   \left(
   1 + {{\cal  M}^{2} \over 4 \kappa^{2}} \right)
\end{equation}
where
\begin{equation}\label{luma37} 
  {\cal M}^{2} (p) =
  \left( 2 \kappa  \sinh \, {p_{0} \over 2 \kappa }\right)^{2} -
  e^{{P_{0}\over \kappa}} \overrightarrow{p}^{2}
\end{equation}
describes the $\kappa$-deformed mass Casimir in bicrossproduct
basis, we obtain  the following counterpart of the
$\kappa$-deformed KG equation  on standard Minkowski space
\begin{equation}\label{luma38}      
  {\cal M}^{2} \left( {1\over i} \partial_{\mu} \right)
  \left( 1 + { {\cal M}^{2}( {1\over i} \partial_{\mu} )\over
  4\kappa^{2}} \right) \varphi(y) = m^{2}_{0} \, \varphi(y)
\end{equation}
We see that the $\kappa$-deformed KG operator
     contains additional
tachyon of the mass  $-2 \kappa^2 (1+(1+m^2/
\kappa^2)^{\frac{1}{2}}) $.

(b) Dirac equation.

On $\kappa$-deformed Minkowski space, due to lack of deformation
in Lorentz sector the algebra of Dirac matrices is not deformed,
i.e. we get
\begin{equation}\label{luma39}  
  \left\{ \gamma^{\mu},\gamma^{\nu} \right\} = 2\eta^{\mu\nu}
\end{equation}
The Dirac equation on $\kappa$-deformed Minkowski space
\begin{equation}\label{luma40} 
  \left(
 \gamma^{\mu} \widehat{\partial}_{\mu} - m_{0} \right)_{AB}
 \, \Psi_{B} ( \widehat{x} ) = 0
\end{equation}
has the following counterpart on standard Minkowski space (see
also (\ref{luma34}))
\begin{eqnarray} \label{luma41}  
&&  \left[ \gamma^{0}
\left(
\kappa \,
   \sin \, {\partial_{0} \over \kappa}
-
  {1\over 2 \kappa} \,
e^{i {\partial_{0}\over \kappa}} \Delta
  \right) \right.
  \cr
  &&
\left.
   -
  e^{i {\partial_{0}\over \kappa}}
  \gamma^{i}      \partial_{0} +
  m_{0} \right]_{AB} \, \Psi_{B} = 0
\end{eqnarray}

The square of the $\kappa$-deformed Dirac operator provides the
$\kappa$-deformed K.G. operator (\ref{luma37}).

The equation (\ref{luma41}) has been firstly given in    [19,20]

(c) Proca and Maxwell equation.

The Proca equation on $\kappa$-deformed Minkowski space takes the
form:
\begin{equation}\label{luma42}  
  \left( \widehat{\partial}_{\mu} \widehat{\partial}^{\mu}
  \delta_{\nu} ^{\ \rho} -
\widehat{\partial}_{\nu}
\widehat{\partial}^{\rho} \right)
  U_{\rho} ( \widehat{x}) = m^{2}_{0} U_{\rho} ( \widehat{x})
\end{equation}
or equivalently (we observe that $\left[ \widehat{\partial}_{\mu},
\widehat{\partial}_{\nu} \right] = 0$)
\begin{equation}\label{luma43}  
   \widehat{\partial }^{\mu} \, F_{\mu\nu}( \widehat{x}) =
  m^{2}_{0} \, U_{\nu} ( \widehat{x})
\end{equation}
where
\begin{equation}\label{luma44}    
  F_{\mu\nu}( \widehat{x}) = \widehat{\partial}_{\mu} U_{\nu} (x)
   - \widehat{\partial}_{\nu} U_{\mu} (x)
\end{equation}
If $m=0$ we get
\begin{equation}\label{luma45}  
  \widehat{\partial}^{\mu} F_{\mu\nu} ( \widehat{x}) = 0
\end{equation}
and the theory is invariant under the following local $U(1)$
gauge transformations (we denote $U_{\mu}\big|_{m_{0}=0} =
A_{\mu}$):
\begin{equation}\label{luma46}  
  A^{\prime}_{\mu} ( \widehat{x}) = A_{\mu} ( \widehat{x}) -
  \widehat{\partial}_{\mu} \, \alpha ( \widehat{x})
\end{equation}
The formula (\ref{luma46})
 has the following counterpart on commutative
Minkowski space:
\begin{eqnarray}\label{luma47}  
   A^{\prime}_{i}\left( \overrightarrow{y}, y_{0} \right)
=
  A_{i} \left(\overrightarrow{y},y_{0} \right) -
  \partial_{i} \, \alpha
\left( \overrightarrow{y},
   y_{0} - {i\over \kappa}
     \right) &&
  \cr
 A^{\prime}_{0} \left( \overrightarrow{y}, y_{0} \right)
=
  A_{0}  \left( \overrightarrow{y},y_{0} \right) -
    \kappa \left(
   \alpha
   \left( \overrightarrow{y}, y_{0}
   +
   {i\over  \kappa}
    \right)     \right. &&
\cr
 -
     \alpha \left(
 \overrightarrow{y},
   y_{0} - {i\over \kappa}
    \right)
    +
    {1\over 2 \kappa} \, \Delta \, \alpha
\left(
     \overrightarrow{y}, y_{0} + {i\over \kappa}
\right)
&&
\end{eqnarray}
One can calculate that the product of two $\kappa$-deformed
$U(1)$ gauge transformations is given by the formula:
\begin{equation}\label{luno48}    
:\,   e^{i\alpha ( \widehat{x})}\, : \, : \,
  e^{i\alpha^{\prime}( \widehat{x})} =
  e^{i\alpha^{\prime\prime}_{\kappa}( \widehat{x})}  \, :
\end{equation}
where
\begin{eqnarray}\label{luma50x}
  \widehat{\alpha}^{\prime\prime}_{\kappa}
  (\overrightarrow{\widehat{x}},\widehat{x}_{0}) &=&
  \alpha^{\prime}
  (\overrightarrow{\widehat{x}},\widehat{x}_{0})
  \cr
  &&  + \,
  \alpha \left(
  \exp i \left[
  \alpha^{\prime} \left(
  (\overrightarrow{\widehat{x}},\widehat{x}_{0} - {i\over \kappa}
  \right) \right.\right.
  \cr
  &&
  -
\left.\left.
  \alpha^{\prime} \left(
  \overrightarrow{\widehat{x}},\widehat{x}_{0}\right)\right]
\overrightarrow{\widehat{x}},\widehat{x}_{0}\right)
\end{eqnarray}

\section{On the
  second quantization of $\kappa$-deformed free fields}

Let us consider as an example the second quantization of
$\kappa$-deformed KG field
 (see (40))
using the commutative
 space-time framework. The relativistic wave equation
 (\ref{luma38}) can be written
 as the following product of standard and tachyonic deformed
 KG factors:
\begin{eqnarray}\label{luma51}
&&  \left( {\cal M}^{2} \left( {1\over i}\partial_{\mu}\right) -
  m^{2}_{+} \right)
  \cr
&& \qquad
\cdot \left( {\cal M}^{2} \left( {1\over
i}\partial_{\mu}\right) -
  m^{2}_{-} \right)
  \varphi(x) = 0
\end{eqnarray}
where
\begin{eqnarray}\label{luma52}
  m^{2}_{+} & = & m^{2}_{0} -  {m^{4}_{0}\over 4\kappa^{2}} +
  \mathcal{O}
  \left( {1\over \kappa^{4}}\right)
\cr
 m^{2}_{-} & = & -  4\kappa^{2}
  - m^{2}_{0} +  {m^{4}_{0}\over 4\kappa^{2}} + \mathcal{O}
  \left( {1\over \kappa^{4}}\right)
\end{eqnarray}
We see that $\varphi(x)$ is the linear combination of two
solutions $\varphi_{+}, \varphi_{-}$ satisfying the equation
\begin{equation}\label{luma53}
 \left( {\cal M}^{2} \left( {1\over i}\partial_{\mu}\right)
 - m^{2}_{\pm} \right) \varphi_{\pm}(x) = 0
\end{equation}
The physical solution of (\ref{luma53}) with $m^{2}_{+} > 0$ can
be quantized. Let us write
\begin{eqnarray}\label{luma54}
  \widehat{\varphi}_{\pm}(x)
 & =&  {1\over (2\pi)^{2}}
  \int d^{4}\, p \delta
\left( {\cal M}^{2}(p)
 - m^{2}_{+} \right)
 \cr
  && \qquad \qquad \qquad  \qquad \cdot  \varphi_{+}(p)e^{ipx}
\cr
 &= & {1\over (2\pi)^{2}} \int
 { d^{3} \vec{p}\over
 \omega_{\kappa}(\vec{p}) }
 \cr
 && \qquad        \cdot
 \left( a ( \vec{p}) e^{i
 \vec{p}\vec{x} - \omega_{\kappa}
( \vec{p}) t} + H.C. \right)
\end{eqnarray}
where
 $p_{0}=\omega(\vec{p})$
 describes the dispersion relation for
$\kappa$-deformed relativistic particle (see (39)).
As the first attempt let us introduce the
following $\kappa$-deformed creation and anihilation relations
\begin{equation}\label{luma55}
  \left[ a^{+}(\vec{p}), a(\vec{p}^{\ \prime})\right]
  = 2\omega_{\kappa}(p) \delta^{3}
  (\vec{p} - \vec{p}^{\ \prime} )\, .
\end{equation}
If we define the Fock vacuum
as follows
\begin{equation}\label{luma56}
  a^{+}(p) | 0\rangle = 0
\end{equation}
one gets
\begin{eqnarray}\label{luma57}
&&  \langle 0 | \widehat{\varphi}_{\pm} (x),
  \widehat{\varphi}_{\pm} (x^{\prime}) |0 \rangle  =
\\
&&
  =
  {1\over (2\pi)^{4}}
  \int {d^{3} \vec{p}\over 2\omega_{\kappa}(\vec{p})}
e^{i\omega_{\kappa} (\vec{p})(t-t^{\prime})}
  \cr
 && = {1\over (2\pi)^{4}}
\int d^{4} p \theta(p_{0})\delta
\left( {\cal M}^{2}(p)
 - m^{2}_{+} \right)
 e^{ip(x-x^{\prime})}
\nonumber
\end{eqnarray}
We see that the formula (\ref{luma57}) describes the
$\kappa$-deformation of free scalar Wightmann function. It should be
mentioned here that the $\kappa$-deformed Green functions in
standard basis, obtained by the redefinition $p_{i}\to e^{ -
{P_{0}\over \kappa}} p_{i}$, were  discussed in [21].

It should be stressed, however, that the relation (56) is not
consistent with the $\kappa$-covariant definition of two-particle state
$| p_{1};p_{2}\rangle$
 introduced as
$| p_{1};p_{2}\rangle = a (p_{1})a(p_{2})|0\rangle$.
It is easy to see that the relation
$|p_{1};p_{2}\rangle=|p_{2},p_{1}\rangle$
 is not consistent with
  the nonsymmetric coproduct $\Delta^{(2)}(\overrightarrow{p})$ describing
   the threemomenta of two-particle states.
We get
\renewcommand{\theequation}{59\alph{equation}}
\setcounter{equation}{0}
    \begin{eqnarray}
    \overrightarrow{P} | \overrightarrow{p}_{1},E_{1};
    \overrightarrow{p}_{2},E_{2}\rangle
&=& \left(
 \overrightarrow{p}_{1}\, e^{{ E_{2}\over \kappa\, c}}
+  \overrightarrow{p}_{2}
 \right)
\cr
&&
|
 \overrightarrow{p}_{1}, E_{1}; \overrightarrow{p}_{2},E_{2}
\rangle
\end{eqnarray}
 \begin{eqnarray}
    \overrightarrow{P}
|\overrightarrow{p}_{2},E_{2};
    \overrightarrow{p}_{1},E_{1}
\rangle
&=& \left(
\overrightarrow{p}_{2}\, e^{{ E_{1}\over \kappa\, c}}
+ \overrightarrow{p}_{1}
   \right)
\cr
&&
|
\overrightarrow{p}_{1},E_{1};\overrightarrow{p}_{2}
\overrightarrow{E}_{2}
\rangle \, .
\end{eqnarray}
\renewcommand{\theequation}{\arabic{equation}}
\setcounter{equation}{59}
It is easy to see that
both  formulae (59a-b) give the same value of
total two-particle
 threemomentum if we assume
\begin{eqnarray}\label{luma60}
  a\left(
  \overrightarrow{p}_{1},E_{1}\right)
  a\left(\overrightarrow{p}_{2},E_{2}\right)
&  = & a\left( \overrightarrow{p}_{2}\,
  e^{ - {E_{1}\over \kappa \, c}}, E_{2}\right)
\\
&&
a\left( \overrightarrow{p}_{1}\,
  e^{ - {E_{2}\over \kappa \, c}}, E_{1}\right)\, .
\nonumber
\end{eqnarray}
The flip operation (60) can be represented by the action of
quantum $R$-matrix,
which for the quantum $\kappa$-Poincar\'{e} algebra
is not known.

 Further considerations leading to the $\kappa$-Poincar\'{e}
covariant Fock space and corresponding formulation of
 second-quantized fields on noncommutative Minkowski space
 is under investigation.

\section{Conclusions}
Let us summarize our results. Firstly we have defined the
covariant
 wave functions on
$\kappa$-deformed Minkowski space $M_{\kappa}$.  Their
transformation properties are defined by the right coacting of
$P_{-\kappa}$\ - $\kappa$-Poincar\'{e} group with the parameter
$\kappa$\ replaced by $-\kappa$.  We solved the problem of
finding the intertwinners between this representation acting on
coordinate wave functions and the unitary representation of
$P_{\kappa}$\ acting in momentum space. These  intertwinners are
actually the wave functions of particles of definite mass and
spin.
 The main result here is that once this problem is solved in
the nondeformed case it can be similarly solved also for
nonvanishing deformation parameter.

In second  part of this      report we define the basic deformed
free fields on noncommutative Minkowski as well as on standard
Minkowski space. Due to star product technique one can represent
the effect of noncommutativity of $ \kappa$-Minkowski space by
nonlocal deformation of the derivatives defining free field
equations (see (\ref{luma34})). It should be stressed that
$\kappa$-deformation implies
 the deformation of free equations and
 the nonlocality in time.

In order to     define the $\kappa$-deformed second-quantized
field theory one should quantize the Fourier transforms $a(q)$
(see (\ref{wz20})) or $\widetilde{a}_{A}(p)$ (see
(\ref{luma29})). This problem was
briefly discussed in Sect. 5 and requires for its further understanding
 the $\kappa$-covariant structure of multiparticle states
 and consistency with nonAbelian addition law for
 $\kappa$-deformed threemomenta.

\end{document}